\documentclass[aps,pre,a4paper,10pt,twocolumn,showkeys,reprint,floatfix]{revtex4-1}
\usepackage{amsmath}
\usepackage{amssymb}
\usepackage{graphicx}
\usepackage{csquotes}
\usepackage[colorlinks=true,hyperfootnotes=true,breaklinks=true]{hyperref}

\begin{document}

\title{Ageing of complex networks}

\author{Zdzislaw~Burda}
\thanks{\includegraphics[scale=0.1]{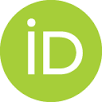}~\href{https://orcid.org/0000-0002-9656-9570}{0000-0002-9656-9570}}
\email{zdzislaw.burda@agh.edu.pl}

\author{Michalina~Kotwica}

\author{Krzysztof~Malarz}
\thanks{\includegraphics[scale=0.1]{ORCID.png}~\href{https://orcid.org/0000-0001-9980-0363}{0000-0001-9980-0363}}
\email{malarz@agh.edu.pl}

\affiliation{\href{http://www.agh.edu.pl/}{AGH University of Science and Technology, Faculty of Physics and Applied Computer Science,
al. Mickiewicza 30, 30-059 Krakow, Poland.}}


\begin{abstract}
Many real-world complex networks arise as a result of a competition between growth and rewiring processes. Usually the initial part of the evolution is dominated by growth while the later one rather by rewiring. The initial growth allows the network to reach a certain size while rewiring to optimise its function and topology. As a model example we consider tree networks which first grow in a stochastic process of node attachment and then age in a stochastic process of local topology changes. The ageing is implemented as a Markov process that preserves the node-degree distribution. We quantify differences between the initial and aged network topologies and study the dynamics of the evolution. We implement two versions of the ageing dynamics. One is based on reshuffling of leaves and the other on reshuffling of branches. The latter one generates much faster ageing due to non-local nature of changes.
\end{abstract} 

\maketitle

\section{Introduction}

Random graphs have continuously attracted attention of researchers in mathematics, 
physics, computer science and many other research fields since the fifties of the last century \cite{Erdos_1959,Gilbert_1959}. In the nineties, due to advances in computer technologies and data mining, researches were able to collect and systematically analyse enormous empirical data sets on technological networks, real world networks and abstract networks used to describe complex systems \cite{Barabasi_1999,Albert_2002,Dorogovtsev_2002,Newman_2003}. The analysis led to the
understanding of network topology, its structure and functionality. It also led to a landmark discovery of principles underlying the emergence of scaling laws and highly heterogeneous architecture of real-world complex networks, including the Internet \cite{Barabasi_1999, Dorogovtsev_2000, Krapivsky_2000, Krapivsky_2001}. This paved the way for new ideas and models that aimed to explain observed features of complex networks. The models drew inspiration from statistical physics, combinatorics, graph theory and computer science and in the course of time they have evolved into a scientific discipline which is known as complex network science today. Complex network science finds applications in many research areas ranging from genetics \cite{Barabasi_2004}, epidemiology \cite{Colizza_2015}, 
ecology \cite{Sole_2001}, through linguistics \cite{Motter_2002}, 
economy \cite{Onnela_2003}, sociology \cite{Fortunato_2009}, computer science \cite{Yan_2006}, physics \cite{Dorogovtsev_2008} to telecommunication \cite{Onnela_2007}, transportation \cite{Helbing_2001} and many others.

The main focus of complex network science in its early days was on growing networks. By adapting the Yule process \cite{Yule_1924,Simon_1955} one was able to explain the 
scale-free tails of node-degree distribution and the heterogeneity of network architecture observed in many empirical data sets
\cite{Barabasi_1999, Dorogovtsev_2000, Krapivsky_2000, Krapivsky_2001}. At the same time a statistical approach was developed \cite{Burda_2001,Berg_2002,Palla_2004}. In this approach complex networks are viewed as random graph ensembles equipped with a probability measure. Using the probability measure one can define entropy of random graphs, determine physical quantities as ensemble averages, address the question of fluctuations and self-averaging, and study relaxation and thermalisation processes.

{In many real-world situations networks emerge as a result of a growth process. For example technological networks like the Internet, the World Wide Web, power-grid networks, telecommunication networks, airline transportation networks have grown from scratch. 
The same holds for social networks like the LinkedIn network or a network of phone numbers stored in mobile phones, as well as for biological networks like gene regulatory networks or for networks in economics like those used to describe linkages between financial institutions, etc. For some networks the growth continues forever, for example for the citation or collaboration networks, but for some the growth may slow down or terminate completely at some stage of evolution when the network reaches a certain size. In this case also other processes may become important. They can be called adaptation or thermalisation processes. They may enforce local modifications of network topology like rewiring of edges, as well as adding or removing edges, nodes or other network motifs. 
 For example: airlines may change connections to optimise market targets; power-grids are continuously restructured by replacing old generation stations by new ones, and by adding/removing transmission lines to improve the efficiency and to adapt to the demand; many social networks like friendship networks keep on changing all the time; gene regulatory networks adapt their topology to optimise the vital functions of the organism they represent, etc. In all these cases growth processes dominate at the beginning of the network evolution being responsible for connecting entities/nodes \cite{Barabasi_1999, Albert_2002, Dorogovtsev_2002,Newman_2003, Dorogovtsev_2000, Krapivsky_2000, Krapivsky_2001}, while later also adaptation mechanisms and thermalisation processes become important for shaping the network architecture,  
\cite{Watts_1998, Krapivsky_2001, Burda_2001, Maslov_2002, Burda_2011}. Sometimes they become dominant. We consider such a situation in this work. For the sake of simplicity, we propose a simple model of networks which initially grow to a certain size and then age by adaptation or thermalisation processes which preserve the size of the network. We are interested in the properties of the ageing process.}

In this paper we discuss random trees which is the simplest class of random networks. It is an important class because there are many exact, explicit, analytical results on random trees \cite{Drmota_2009}. For this reason they often serve as a testing ground for new ideas and algorithms in network science. In this paper we use trees to study ageing phenomena for complex networks. We consider a process which consists of two phases. During the initial phase trees grow to a certain size. Once they reach the size  they start to evolve under the dynamics which preserves the size and the node degree distribution. It brings the initial trees to a stationary state which maximises entropy under the condition that trees have asymptotically the same node-degree distribution as the initial growing trees. The evolution is realised as a repetitive process of cutting a randomly selected leaf at pasting it at a random node of the tree in a way that fulfils the detailed balance condition. This is implemented by the Metropolis--Hastings algorithm. Moving a leaf from place to place changes topology of the tree locally. This mimics the rewiring operation known from Monte Carlo simulations of simple graphs \cite{Watts_1998, Krapivsky_2000, Krapivsky_2001, Burda_2001, Maslov_2002}. We discuss the dynamics of the evolution and compare it to the evolution driven by non-local transformations where whole branches of the tree are cut and moved. As we shall see the non-local version of the algorithm significantly reduces autocorrelations of trees generated during the evolution and speeds up the ageing process. 

The paper is organised as follows. First we recall the construction of growing trees by a stochastic node attachment, with the uniform and preferential attachment kernels, which generate trees with exponential and scale-free node degree distributions, respectively. Next we discuss the Markov Chain Monte Carlo (MCMC) method which can be used to sample  equilibrium trees. The MCMC method will be then used as an ageing process which brings the growing trees to the state of maximal entropy with the same limiting node degree distribution as the initial trees. We compare statistical properties of the initial trees and the aged trees. In particular we compare the node-to-node distance distribution and the branch-size distribution as well as the tree-crown distribution for these trees. Finally we discuss dynamical features of the evolution, in particular we compare the autocorrelation time for evolution based on local or non-local transformations of tree topology which are implemented by reshuffling of leaves or reshuffling of branches, respectively. In Appendix~\ref{app} we discuss an analytic map between the partition functions for the model of weighted trees and the model of weighted partitions. This map is used to derive analytic expressions, for instance for the node-degree distribution for weighted trees. We conclude the paper with a brief summary.

\section{Growing random trees}

Growing trees are constructed by a recursive node attachment \cite{Barabasi_1999, Dorogovtsev_2000, Krapivsky_2000, Krapivsky_2001}. In a single step a new node is attached with a new edge to a randomly selected node of the tree. The number of nodes (and of edges) increases by one. This process is repeated until a desired size is reached. The simplest case is a uniform attachment where nodes are selected uniformly, with the probability $1/n$, where $n$ is the number of nodes of the current tree. The node degree distribution $\pi_n(q)$ approaches  a limiting exponential law $\pi_n(q) \to \pi(q)$   for $n\to \infty$
\begin{equation}
\pi(q) = 2^{-q}, \quad q=1,2,\ldots .
\label{exp_grow}
\end{equation}
Another interesting example is a preferential attachment 
\cite{Yule_1924,Simon_1955}. In this case a node to which the new node is attached is selected with the probability proportional to its degree. This process leads to the following limiting node degree distribution \cite{Barabasi_1999, Dorogovtsev_2000, Krapivsky_2000, Krapivsky_2001}
\begin{equation}
\pi(q) =  \frac{4}{q(q+1)(q+2)}, \quad q=1, 2, \ldots .
\label{ba_grow}
\end{equation}
For large $q$ the distribution asymptotically behaves as a power law $\pi(q) \sim q^{-3}$. For large but finite $n$, the distribution $\pi_n(q)$ slightly deviates from the limiting one (see Appendix~\ref{app}). For example it does not extend to infinite $q$'s but has a clear cut-off behaviour. The shape of the finite size corrections can be determined analytically \cite{Waclaw_2007}. The distribution (\ref{ba_grow}) has an infinite variance, which has very profound consequences for network topology, the main of which is the occurrence of hubs that is nodes of high degree. 

Here we are interested in the ensemble of trees of size $n$. The ensemble can be obtained by repeating the growth process. Each time the process can be initiated from a single node and terminated when the tree has $n$ nodes. The process can be repeated as many times as needed in order to get a sufficiently large sample and thus to estimate physical quantities with a desired accuracy.

\section{Maximal entropy random trees}

The statistical ensemble of maximal entropy random trees on $n$ nodes is defined as an ensemble of equiprobable labelled trees. The partition function is
\begin{equation}
Z_n = \sum_{t\in T_n} 1,
\label{cayley}
\end{equation}
where $t$ runs over the set, $T_n$, of labelled trees on $n$ nodes. There are $n^{n-2}$ trees. Equiprobable trees (\ref{cayley}) are sometimes called free trees or Cayley trees
\cite{Drmota_2009}. The ensemble average of a physical quantity $O$ is defined as 
\begin{equation}
\langle O \rangle_n = \frac{1}{Z_n}\sum_{t\in T_n} O_t.
\end{equation}  
In particular, the node degree distribution for the ensemble is calculated as
\begin{equation}
\pi_n(q) = \left\langle \frac{1}{n}\sum_{i=1}^n \delta_{q_iq} \right\rangle_n,
\end{equation}
where $\delta_{rq}$ is the Kronecker delta. The node-degree distribution $\pi_n(q)$ can be determined analytically (see Appendix~\ref{app}). The limiting distribution $\pi_n(q) \to \pi(q)$ for $n\to \infty$ is 
\begin{equation}
\pi(q) = \frac{1}{e} \frac{1}{(q-1)!}, \quad q=1,2,\ldots
\label{pi_cayley}
\end{equation}
The idea is to slightly weaken the maximal entropy principle and to maximise entropy under condition that trees have a desired node degree distribution. To that end one can consider an ensemble of weighted random trees with the partition function
\begin{equation}
Z_n = \sum_{t\in T_n} W_t =  \sum_{t\in T_n} \prod_{v\in t}^n w(q_v),
\label{weighted}
\end{equation} 
where the statistical weight $W_t=\prod_{v\in t} w(q_v)$ of trees in this ensemble depends only on the node degree sequence. The product is over nodes $v$ of the tree $t$. The node weight $w(q)$ is a non-negative function defined for $q=1,2,\ldots$. It is identical for all nodes. The probability of occurrence of a tree $t$ in the ensemble is 
\begin{equation}
P_t = \frac{W_t}{Z_n} = \frac{1}{Z_n} \prod_{v\in t} w(q_v).
\label{Pt}
\end{equation}
The entropy is maximal in a subclass of trees with the given degree sequence since all trees in this class are equiprobable. The freedom in choosing the weight function $w(q)$ can be used to obtain a tree ensemble with a desired node degree distribution. Let us denote the desired node-degree distribution by $\pi_d(q)$. It is a non-negative function defined on $q=1,2,\ldots$ which is properly normalised $\sum_q \pi_d(q)=1$. The mean must be equal two $\sum_q q \pi_d(q) = 2$ since for trees the mean node degree $2(1-1/n) \to 2$ for $n\to \infty$. Choosing the weight function in (\ref{weighted})
\begin{equation}
w(q) = (q-1)! \pi_d(q)
\label{w_pi}
\end{equation}
one obtains random trees with the desired limiting node-degree distribution $\pi_n(q) \to \pi_d(q)$ for $n\to \infty$, as shown in Appendix~\ref{app}. In particular, for
\begin{equation}
w(q) = 2^{-q }(q-1)! 
\label{w_exp}
\end{equation}
the limiting degree distribution is equal to (\ref{exp_grow}) and for
\begin{equation}
w(q) = \frac{4 (q-1)!}{q(q+1)(q+2)} = \frac{4}{\binom{q+2}{q-1}}
\label{w_ba}
\end{equation}
to the distribution (\ref{ba_grow}), which correspond to exponential and scale-free trees, generated by the uniform and preferential attachment, respectively.
It is worth noting that the partition function (\ref{weighted}) changes by a constant factor under the following transformation of the weight function
\begin{equation}
w(q) \to \tilde{w}(q) = \alpha \beta^q w(q),
\label{alpha_beta_inv}
\end{equation}
so the transformation has no effect on the ensemble averages. Indeed under this change the
partition function transforms as
\begin{equation}
Z_n \to \tilde{Z}_n = \alpha^n \beta^{2(n-1)} Z_n.
\end{equation}
The pre-factor is a constant number for a given $n$, independent of the degree sequences. The invariance under rescaling (\ref{alpha_beta_inv}) tells us for example that skipping the factor $2^{-q}$ in (\ref{w_exp}) or the factor $4$ in (\ref{w_ba}) will have no effect on the ensemble of trees.

So far we have addressed the question how to reproduce a desired limiting node degree distribution by choosing appropriate weights. But one can ask an opposite question: what is the limiting node degree distribution for a given weight function $w(q)$ (\ref{weighted}). We assume only that $w(q)$ is a non-negative function on $q=1,2,\ldots$.   The answer is
\begin{equation}
\pi(q) = \frac{\alpha w(q) \beta^{q} }{(q-1)!},
\label{alpha_beta}
\end{equation}
where the parameters $\alpha$ and $\beta$ are chosen in such a way as to fix the normalisation $\sum_q \pi(q) =1$ and the mean $\sum_q q \pi(q) = 2$. We discuss the derivation of (\ref{alpha_beta}) in Appendix~\ref{app}. To give 
a few examples: 
if $w(q)=1$ then $\alpha=e^{-1}$, $\beta=1$ and $\pi(q)=e^{-1}/(q-1)!$;
if $w(q)=(q-1)!$ then $\alpha=1$, $\beta=1/2$ and $\pi(q) = 2^{-q}$;
if $w(q)=q!$ then $\alpha=4/3$, $\beta=1/3$ and $\pi(q) = 4 q 3^{-(q+1)}$;
if $w(q)=(q-1)!/(q(q+1)(q+2))$ then $\alpha=4$, $\beta=1$ and $\pi(q)=4/(q(q+1)(q+2))$.
As a final remark we note that for some weight functions $w(q)$ there are no such constants $\alpha$ and $\beta$ (\ref{alpha_beta}) that would fix the normalisation $\sum_q \pi(q) =1$ and the mean $\sum_q q \pi(q) = 2$. In this case the corresponding trees collapse to a bush structure characterised by the occurrence of a single vertex of order ${\cal O}(n)$ 
\cite{Bialas_1996, Bialas_1997,Jonsson_2011}. 

\section{Monte Carlo simulations}

Random growing trees can be directly sampled by repeating the growth process
many times from $1$ to $n$ nodes. Trees generated in this way form an independent sample of trees on $n$ nodes which can be used to estimate ensemble averages by sample means. The method is efficient since the acceptance rate is one hundred percent and trees are independent of each other. 

Free random trees (\ref{cayley}) can also be directly sampled
using a bijective map between labelled trees and their Pr\"ufer codes. We briefly mention this construction in Appendix~\ref{app} \cite{Pruefer_1918,Bryant_2015}.

There is no general algorithm to directly sample weighted trees with the probability proportional to the statistical weight (\ref{Pt}). In this case one can apply the Markov Chain Monte Carlo (MCMC) method. The idea is to iteratively generate trees from one another by small modifications called transitions between states. In the MCMC terminology, the consecutive configurations, in our case---consecutive trees, are called states. The consecutive states form a Markov chain $t_0 \to t_1 \to t_2 \to \cdots \to t_N$. Here we shall use the Metropolis--Hastings algorithm \cite{Metropolis_1953,Hastings_1970} which is the best known version of the MCMC method. Let $t$ and $s$ be two states (trees) and let $P(t\to s)$ be the probability that the state $t$ changes to $s$ in a single step of the Markov chain. The transition probability in the Metropolis--Hastings algorithm is
\begin{equation}
P(t\to s) = \min\left\{1, \frac{P_s}{P_t}\right\}.
\label{mh}
\end{equation}
Assume that $t_k=t$ at time $k$. In order to determine the next state $t_{k+1}$ at time $k+1$ a candidate $s$ is uniformly selected and accepted with the probability (\ref{mh}). If the candidate is accepted then we set $t_{k+1}=s$, otherwise $t_{k+1}=t$. The transition probability fulfils the detailed balance condition $P_t P(t\to s) = P_s P(s\to t)$. It is known from general considerations that the detailed balance principle is a sufficient condition for an ergodic Markov chain to generate states $t_0 \to t_1 \to t_2 \to \cdots \to t_N$ with a frequency that approaches $P_t$ for $N\to \infty$. A Markov chain is ergodic if any state can be reached from any other state in a finite number of steps. 
The price to pay is that consecutive states are correlated. The correlations decrease with the distance in the sequence. A tree $t_{n+k}$ obtained from $t_n$ in $k$ Monte Carlo transitions becomes less correlated with $t_n$ when $k$ increases. Since the frequency approaches the probability measure (\ref{Pt}) one can use the trees generated in the Markov chain to estimate ensemble averages of physical quantities on weighted ensemble of random trees. Let $O(t)$ be such a physical quantity, and let $O_i = O(t_i)$ be the value of this quantity on $t_i$. The sample mean $\bar{O} = \sum_i^N O_i/N$ approaches the ensemble mean 
$\bar{O} \to \langle O \rangle$ for $N\to \infty$. The quantities $O_i$ and $O_{i+n}$ are correlated and this has an effect on the broadening of the statistical uncertainty of the sample mean
\begin{equation}
	\sigma = \sqrt{2\tau_{\text{in}} + 1} \sqrt{\frac{\sum_{i=1}^N (O_i-\bar{O})^2}{N(N-1)}}
\label{broadening}
\end{equation}
by a factor $\sqrt{2\tau_{\text{in}} + 1}$ where $\tau_{\text{in}}$ is the integrated autocorrelation time for $O$ \cite{Madras_1988}. For an independent sample $\tau_{\text{in}}=0$,  but for a sample generated by a typical MCMC algorithm $\tau_{\text{in}}$ increases with the size of the simulated systems $n$. This means that the length of the sample must be $\sqrt{2\tau_{\text{in}} + 1}$ times longer than the length of the independent sample in order to obtain a comparable statistical error. It requires an increasing computational resources to achieve a desired accuracy for large systems if the autocorrelation time  $\tau_{\text{in}} = \tau_{\text{in}}(n)$ increases with the system size $n$. We shall discuss this later.

\begin{figure}
\includegraphics[width=0.45\textwidth]{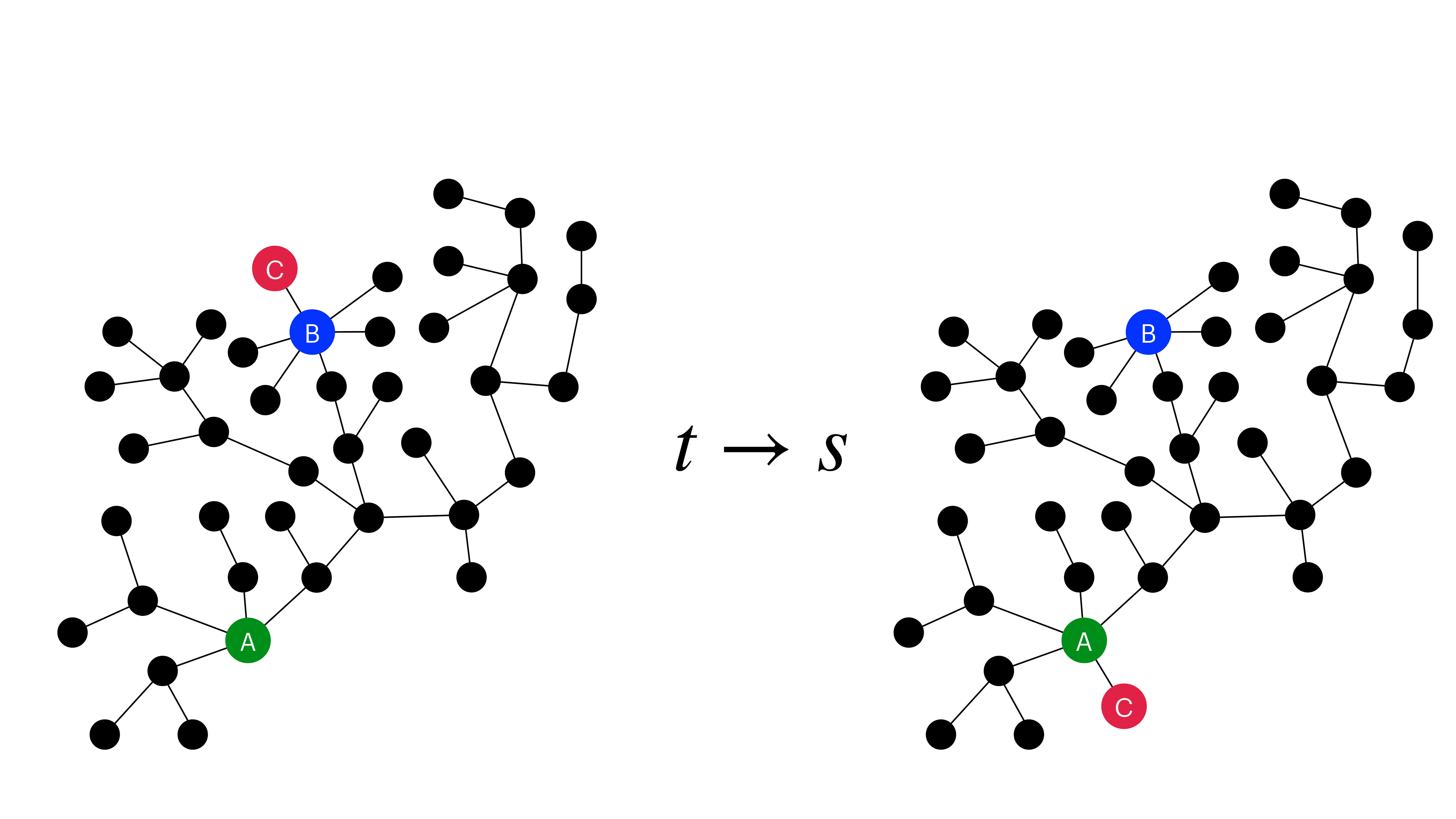}
\caption{(Colour online). An elementary update step of the leaf reshuffling: a leaf (red C) is cut from the vertex it is attached to (marked as blue B) and pasted to a randomly selected vertex (green A). As a result, the degree of the blue vertex decreases by one and of the green vertex increases by one.\label{fig_move_leaf}}
\end{figure}

The Metropolis--Hastings algorithm for weighted random trees (\ref{weighted}) can be implemented by leaf reshuffling which works as follows. A leaf is picked up at random on the current tree, $t$, and moved to a new position at a randomly selected node with the probability (\ref{mh}). As a result, the new tree $s$, which is obtained from $t$, may differ by the position of this single leaf, as illustrated in Fig. \ref{fig_move_leaf}.

Denote the vertex from which the leaf is cut off by $a$ and the one to which it is pasted by $b$. The degrees of the two nodes change by one $q_a \to q_a-1$ and $q_b\to q_b+1$ when the leaf is moved. All others degrees remain unchanged. In the Metropolis--Hastings algorithm, the transition $t\to s$ is accepted with the probability
\begin{equation}
P(t\to s) = 
\min\left\{1, \frac{w(q_a-1) w(q_b+1)}{w(q_a) w(q_b)}\right\} 
\end{equation}
as follows from inserting (\ref{Pt}) to (\ref{mh}). In particular for 
random trees with the exponential limiting node degree distribution (\ref{exp_grow}) the transition probability is
\begin{equation}
P(t\to s) = 
\min\left\{1, \frac{q_b}{q_a-1}\right\},
\label{pts_exp}
\end{equation}
while for random trees with the Yule--Simon limiting node degree distribution (\ref{ba_grow}) 
\begin{equation}
P(t\to s) = 
\min\left\{1, \frac{q_b^2}{q_b+3}\frac{q_a+2}{(q_a-1)^2}\right\} 
\label{pts_ba}
\end{equation}
as follows from (\ref{w_exp}) and (\ref{w_ba}), respectively. More generally,
for random trees conditioned to a desired limiting node degree distribution $\pi_d(q)$ the transition probability (\ref{mh}) is
\begin{equation}
P(t\to s) = 
\min\left\{1, \frac{q_b}{q_a-1}\frac{\pi_d(q_a-1) \pi_d(q_b+1)}{\pi_d(q_a) \pi_d(q_b)}\right\} 
\label{pts_pi}
\end{equation}
as follows from (\ref{w_pi}). Last but not least, the unconditional reshuffling of leaves, $P(t\to s)=1$, generates maximally random trees (\ref{cayley}) with the limiting degree distribution (\ref{pi_cayley}). 

\section{Tree ageing}

\begin{figure}
\includegraphics[width=0.45\textwidth]{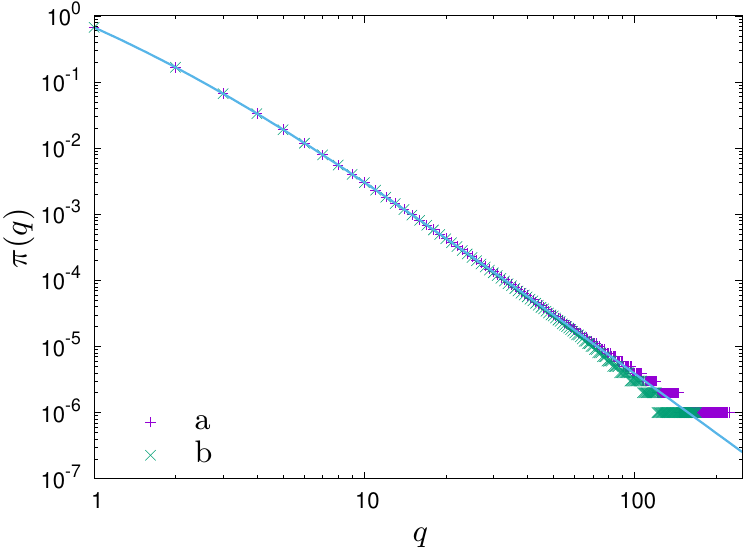}
\caption{
Comparison of the node degree distributions for finite $n$ for the scale-free growing (a) trees and the corresponding aged (b) trees. 
For $n\to \infty$ both the distributions approach the same limiting law given by the Yule--Simon distribution (\ref{ba_grow}). For finite $n$ they differ close to the cut-off.
In the plot we show data for $n=16384$.
\label{fig_ndd_16k_ys}}
\end{figure}

We consider a stochastic process which consists of two phases. The initial phase is the growth by a random node attachment. The growth is terminated once the tree reaches the size of $n$ nodes. The second phase, which is the main part of the evolution, is an ageing process which preserves the size of the tree and the limiting node degree distribution. It is carried out using the Metropolis--Hastings dynamics which brings the initial tree to a stationary state. We address two questions: what are statistical differences between the initial growing trees and the aged ones and what are properties of the ageing process? The stationary state, reached at the end of the ageing process, corresponds to the maximal entropy random trees with the same node degree distribution as the initial trees. Here we discuss exponential trees (\ref{exp_grow}) and scale-free trees (\ref{ba_grow}). In Fig. \ref{fig_ndd_16k_ys} we compare node degree distributions for growing trees obtained by preferential attachment and for the corresponding maximal entropy random trees for finite $n$, obtained as the stationary state of the ageing process.

The limiting node degree distributions for the initial trees and the aged trees become indeed identical when $n\to \infty$ but  for finite $n$ they slightly differ from the limiting ones since they have a finite cut-off induced by finite size of the system. The form of the finite size corrections is slightly different for growing and aged trees \cite{Waclaw_2007,Bialas_1997,Bialas_2000}. We discuss finite size corrections
for aged trees in Appendix~\ref{app}.

\begin{figure}
\includegraphics[width=0.45\textwidth]{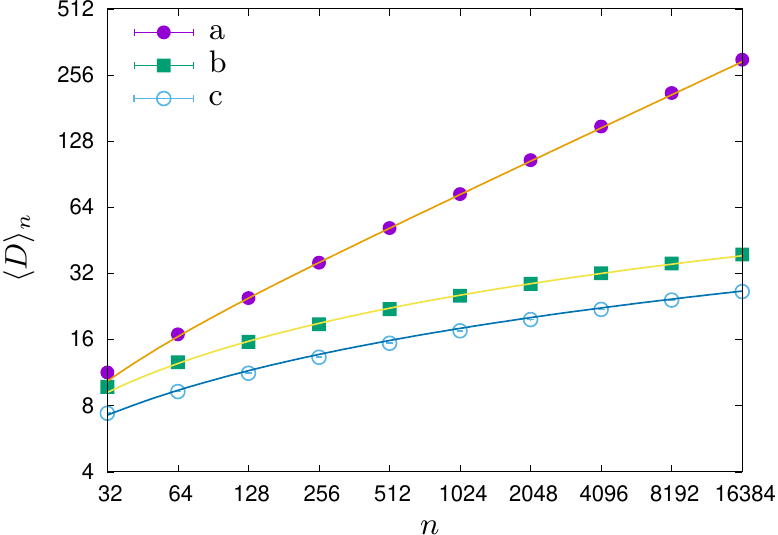}
\caption{The data points show the diameter
	of the aged exponential trees (a), 
	of the exponential growing trees (b), 
	and of the scale-free growing trees (c), 
	for $n=2^5, 2^6, \ldots, 2^{14}$.
	The data points are plotted with error bars. Each point was obtained from the number of measurements of order $10^5$. As a result, the error bars are very small---much smaller than the symbol size. For example for $n=16384=2^{14}$ the diameter of the scale-free growing trees is $26.46(1)$, the diameter of the exponential growing trees is $39.01(1)$ and of the exponential aged trees---$300.5(1)$.
The data for the growing trees is very well described by a logarithmic dependence 
$\langle D\rangle_n = a \log n + b$, 
and for the aged trees by a square root formula with finite size corrections:
$\langle D\rangle_n = a \sqrt{n} (1 + b/n)$. 
The lines shown in the plot correspond to 
$\langle D\rangle_n = 3.1 \log(n)- 3.5$, 
$\langle D\rangle_n = 4.7 \log(n)- 7.1$ and
$\langle D\rangle_n = 2.3 \sqrt{n} (1- 6.6/n)$. 
\label{fig_diameter}}
\end{figure}

The initial trees and the aged ones have the same limiting node degree distribution but they have completely different topology. The most striking difference is that the diameter of random growing trees asymptotically increases as a logarithm $\langle D \rangle_n \sim \log n$ of the number of vertices \cite{Krapivsky_2000, Krapivsky_2001,Bialas_2003} while for the maximally random trees as a square root $\langle D\rangle_n \sim \sqrt{n}$ \cite{Drmota_2009}. Thus, the Hausdorff dimension of random growing trees is infinite 
$d_H=\infty$ while of the corresponding maximally random trees is equal two
$d_H=2$. The average distance between nodes 
\begin{equation}
\bar{d}= \frac{1}{n^2}\sum_{i,j} d_{ij}
\end{equation}
also asymptotically grows as $\langle \bar{d} \rangle_n \sim \log n$ for growing trees 
\cite{Krapivsky_2000, Krapivsky_2001,Malarz_2003,Malarz_2004} and as $\langle \bar{d} \rangle_n \sim \sqrt{n}$ for aged trees \cite{Drmota_2009}. The sum in the last equation is over all pairs $i,j$ of vertices of the tree. We see that typical distances between nodes on the initial trees are much smaller than for the aged ones. In other words the initial trees expand during the ageing process. This holds for both the exponential and scale-free trees. Generally the scale-free trees are less expanded than the exponential ones because of the presence of nodes with large degrees \cite{Krapivsky_2000, Krapivsky_2001}. This is illustrated in Fig. \ref{fig_diameter} where we 
plot the diameter versus $n$ for exponential and scale free initial growing trees and for the exponential aged trees. 

\begin{figure}
\includegraphics[width=0.45\textwidth]{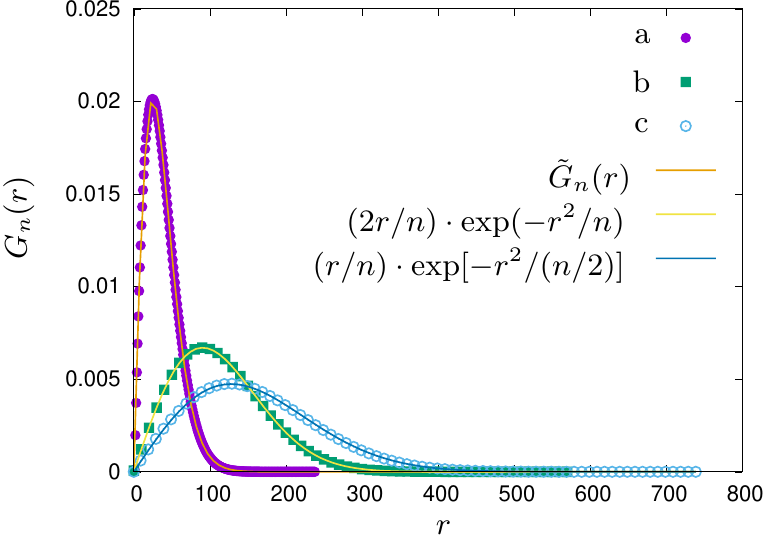}
\caption{From left to right: node-to-node distance distribution for the aged scale-free trees (a), for the aged exponential trees (b) and for the free trees (c) for $n=16384=2^{14}$. Data points are represented as symbols. Histograms are obtained from $10^5$ measurements each. 
Solid lines for the exponential and free trees represent the theoretical expression (\ref{Gn_limiting}) with the parameter $s=2$ and $s=1$ respectively.
These parameters are equal to the variance of the distribution (\ref{exp_grow}) and of (\ref{pi_cayley}).
For the scale free trees we used the phenomenological formula (\ref{G_fs}).
The best fit gives $s=25.81(15)$ and $a=0.0077(13)$.
It very well fits the data.
\label{fig_dd_aged}}
\end{figure}

\begin{figure}
\includegraphics[width=0.45\textwidth]{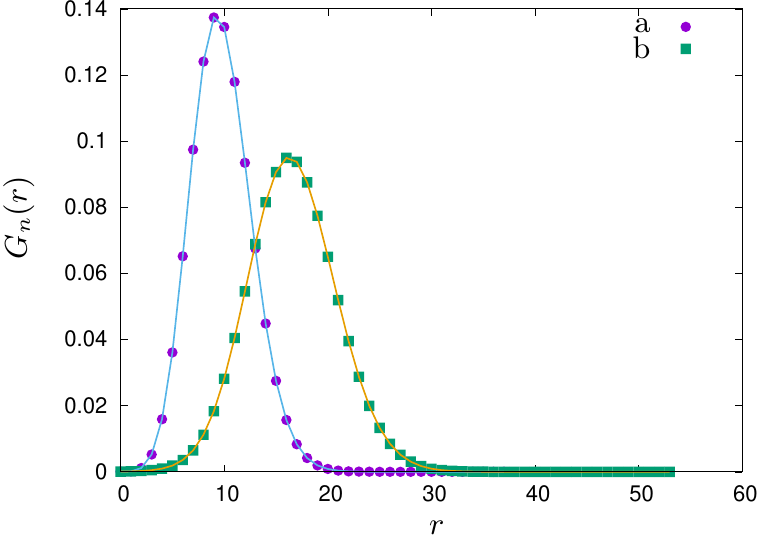}
\caption{Node-to-node distance distributions for growing scale-free trees (a) and growing exponential trees (b) for $n=16384$. The histograms are obtained from measurements on $10^5$ trees. The solid lines between points are drawn to guide the eye.  
\label{fig_dd_grow}}
\end{figure}

The difference between the growing and aged trees is even more clearly seen 
in the distribution of the node-to-node distance. 
This distribution, $G_n(r)$, is defined as the fraction of all pairs of vertices which are in the distance $r$ from each other 
\begin{equation}
G_n(r)= \left\langle \frac{1}{n^2} \sum_{i,j} \delta_{d_{ij}r} \right\rangle_n.
\end{equation}
Clearly $\langle \bar{d}\rangle_n = \sum_r r G_n(r)$. 
For weighted random trees (\ref{weighted}) the node-to-node distance distribution asymptotically approaches a universal limiting shape \cite{Burda_2003}
\begin{equation}
G_n(r) = \frac{s r}{n} \exp\left(-\frac{s r^2}{2n}\right)
\label{Gn_limiting}
\end{equation}
for large $n$, with a single parameter $s$ which is given by the variance of the
node-degree distribution. For the Cayley trees (\ref{pi_cayley}) the variance is $s=1$, for the exponential distribution (\ref{exp_grow}) it is $s=2$. In Fig. \ref{fig_dd_aged} we show data for aged scale-free trees, for aged exponential trees and for Cayley trees for $n=16284$ nodes.
The latter two are compared to the limiting expression (\ref{Gn_limiting}) with $s=2$ and $s=1$, respectively. For the aged scale-free trees the situation is slightly more complicated since in this case the variance of the Yule--Simon distribution (\ref{ba_grow}) is infinite and the limiting formula (\ref{Gn_limiting}) does not hold anymore. In this case we propose a phenomenological approach to derive an approximation for the node-to-node distance distribution for large but finite $n$.
For any finite $n$ there is a finite-size cut-off in the node-degree distribution so the variance $s_n$ exists. We replace $s$ by $s_n$ in (\ref{Gn_limiting}) and additional introduce a finite-size correction by defining an effective distance   
\begin{equation}
R(r) = \frac{r}{\sqrt{1+ a r}},
\end{equation}
where $|a|\ll 1$ is a small parameter. Inserting the effective distance to (\ref{Gn_limiting}) and using the transformation law for the probability distribution
$\tilde{G}_n(r)= R'(r) G_n(R(r))$ we obtain the following finite size expression
\begin{equation}
\tilde{G}_n(r) = \frac{s r}{n} \frac{(1+ a r/2)}{(1+ a r)^2} 
\exp\left( -\frac{s r^2}{2n (1+ a r)}\right).
\label{G_fs}
\end{equation}
For $a=0$ it is of course equivalent to (\ref{Gn_limiting}). 
We used this finite size expression to fit data for aged scale free trees. As one can 
see in Fig. \ref{fig_dd_aged} it indeed very well captures the shape of the curve obtained from the numerical data. All three curves grow linearly $G_n(r) \sim r$ for small $r$ which means that the number of nodes of the tree within the distance $r$ grows quadratically with $r$, as one expects for the fractal dimension equal two. Now let us compare it to the corresponding data for growing trees, Fig. \ref{fig_dd_grow}.

As one can see the node-to-node distance distribution has a completely different shape in this case. The range of the distribution is much shorter than for aged trees, the peak around the maximum is much higher and narrower (compare the scale on the axes in Fig. \ref{fig_dd_aged} and Fig. \ref{fig_dd_grow}). For small $r$ the distribution grows exponentially \cite{Malarz_2004a,Malarz_2006}, and not linearly as before, reflecting the fact that the growing trees have an infinite fractal dimension.

\begin{figure}
\includegraphics[width=0.45\textwidth]{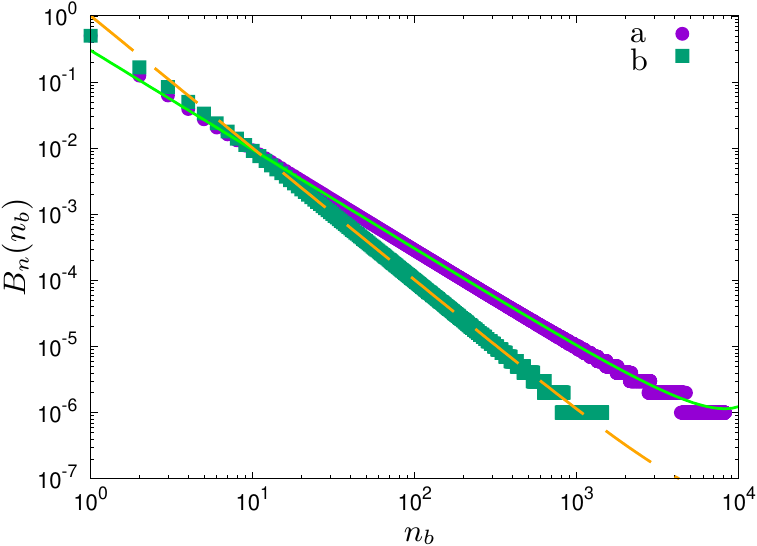}
\caption{The branch size distribution for the exponential growing trees (b), and for the exponential aged trees (a) for $n=16384$. The lines represent the asymptotic expression (\ref{BS}) with $\beta=3/2$ and $\beta=2$, respectively.  
\label{fig_bsd}}
\end{figure}

Another interesting characteristics of tree topology is the branch size distribution. It is a counterpart of the baby-universe distribution known from the studies of random surfaces \cite{Jain_1992}. It is defined as follows. If an edge is cut the tree splits into two subtrees having $n_b$ and $n-n_b$ nodes, the smaller of which, $n_b \ll n-n_b$, is called a branch of the tree. Making a histogram of branch sizes for all edges on a tree one obtains the branch size distribution for this tree. Averaging it over trees one obtains the branch size distribution for the ensemble of trees.  One can analytically determine the asymptotic form of the branch size distribution for free trees (\ref{cayley})
\begin{equation}
B_n(n_b) \sim n_b^{-\beta} \left(1-\frac{n_b}{n}\right)^{-\beta} 
\label{BS}
\end{equation}
for large $n$ and $n_b$. The exponent is $\beta=3/2$ (see Appendix~\ref{app} for details). On the universality grounds one can argue that also for generic weighted trees (\ref{weighted}) the asymptotic distribution has the same form (\ref{BS}) with the same exponent $\beta=3/2$, unless the weights 
(\ref{weighted}) are tuned in a very specific way \cite{Bialas_1996}. In other words we expect that the exponential and scale-free trees obtained by ageing will follow this law. Indeed, the branch size distributions for trees obtained by ageing of the exponential and scale free trees follow, for $n_b\ll 1$, the analytic expression (\ref{BS}) with $\beta=3/2$. This is not any more the case for growing trees for which the exponent changes from $\beta=3/2$ to $\beta=2$, see Fig. \ref{fig_bsd}.

\begin{figure}
\includegraphics[width=0.45\textwidth]{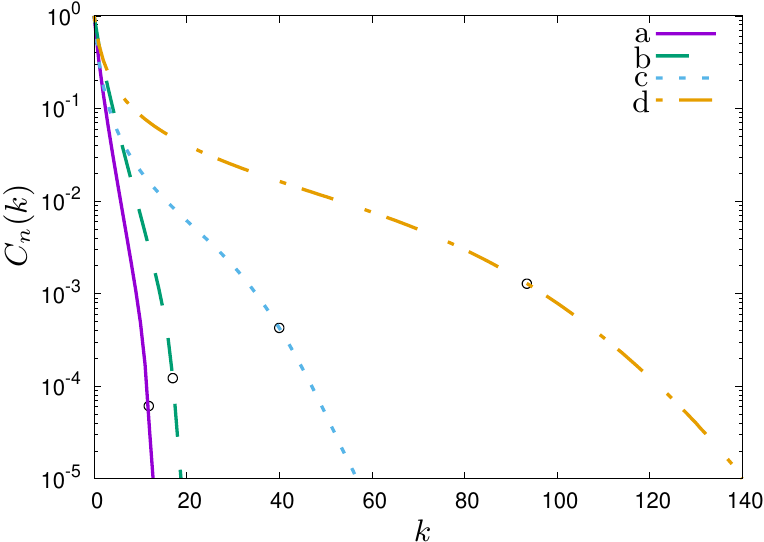}
\caption{From left to right: the crown distribution for the scale-free growing trees (a), the exponential growing trees (b), the scale-free aged trees (c) and the exponential aged trees (d) for $n=16384$. Each histogram was constructed by averaging over $\sim 10^5$ trees. The symbols on the curves mark the average number of pruning steps $\langle K\rangle_n$ at  which the spine is reached: $11.7869(24)$, $17.4636(31)$, $38.965(35)$, $93.543(39)$.
\label{fig_cd}}
\end{figure}

\begin{figure}
\includegraphics[width=0.45\textwidth]{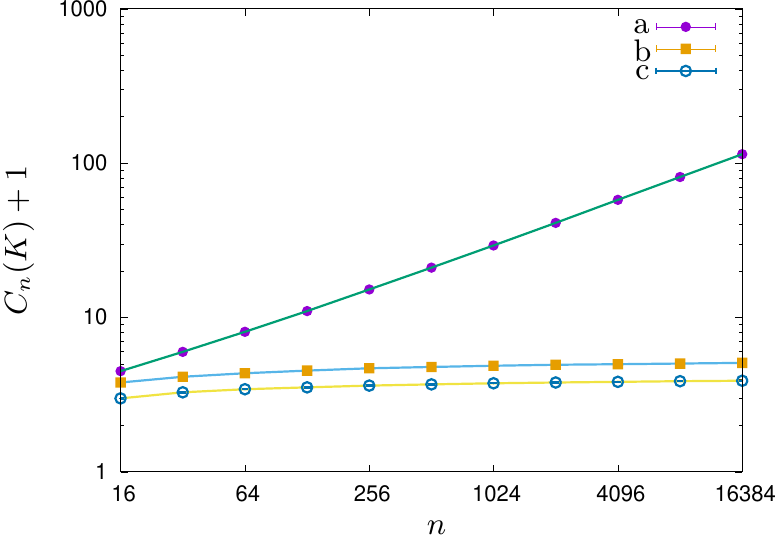}
\caption{\label{fig_stem}The number of nodes of the spine of aged and growing trees. The upper curve represents the data for exponential aged trees (a). It grows like $\sqrt{n}$. The two lower curves show the Monte Carlo data for exponential growing trees (b) and scale-free growing trees (c). The spine length grows very slowly in this case. While $n$ changes from 32 to 16384 the number of nodes in the spine increases roughly from 4 to 5 for the exponential growing trees, and from 3 to 4 for the scale-free growing trees.}
\end{figure}

Another interesting insight into the tree topology is provided by what we call tree--crown distribution. The distribution is obtained by a recursive tree pruning. A tree is pruned by removing all its leaves. The tree remaining after the first pruning can be pruned again and again until it reduces to a linear graph or a single vertex. This linear graph is called spine (or stem) of the tree. Trees obtained by this recursive pruning procedure form a nested set similar to the Matryoshka doll. The tree--crown distribution $C_n(k)$ is defined as the fraction of nodes of the tree after $k$-pruning steps. For $k=0$ it is just $C_n(0)=1$, for $k=1$ it is just the fraction of nodes left after the first pruning step. Clearly, it is equal one minus the percentage of nodes which are leaves of the original tree. For $k=2$ it is the  fraction of nodes left after two consecutive pruning steps, and so on until a naked spine is left. Denote the pruning step at which the spine is reached by $K$. For $k=K$  the crown distribution $C_n(K)$ gives the fraction of vertices which belong to the spine of the tree. For larger $k>K$ the distribution is zero $C_n(k)=0$. For a single tree the distribution has a clear
threshold at $k=K$ where it sharply drops from $C_n(K)$ to zero, but when  the distribution is averaged over many trees, the distribution smooths out and the threshold behaviour is replaced by a smooth cross-over function
which continuously falls off to zero, since the position of the threshold changes from tree to tree. In Fig. \ref{fig_cd}
we compare the crown distribution for different tree ensembles. One can see
that the distribution is much broader for the aged trees than for the corresponding growing trees. One can also compute the average size of the spine.
The result is shown in Fig. \ref{fig_stem}. One can see that the length of the spine weakly depends on the tree size for the growing trees in contrast to the aged trees where it increases as a square root of $n$.

\section{Slow and fast dynamics}

\begin{figure}
\includegraphics[width=0.45\textwidth]{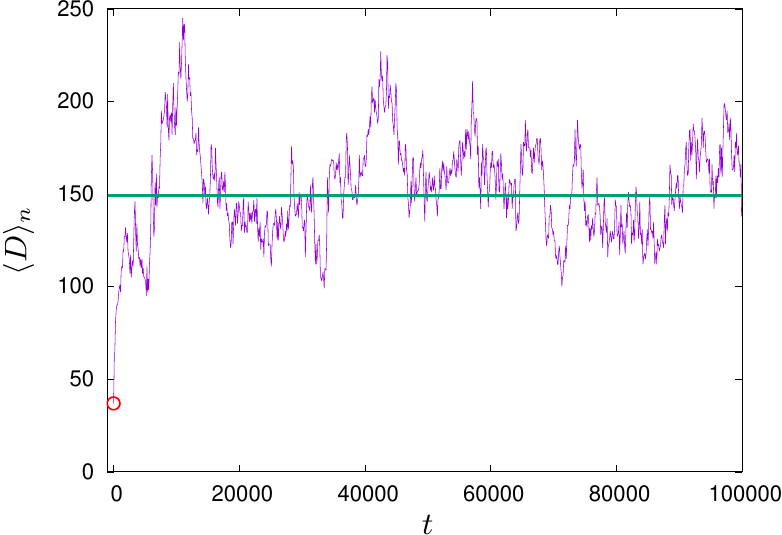}
\caption{The trajectory represents evolution of the diameter of the 
exponential tree initiated from an exponential growing tree for $n=4096$. 
The data points correspond to measurements done every 100th sweep.  The initial value, $37$, 
is marked by the circle and the stationary state value, $149.230(76)$, by the horizontal line. One can see long wave oscillations around the equilibrium value, which reflect large autocorrelations of trees generated by the leaf reshuffling. 
\label{fig_slow}}
\end{figure}

As we have seen in the previous section, the architecture of growing trees is completely different than of the corresponding aged trees. In this section we study ageing dynamics. In particular we are interested in the relaxation time
that is the time needed to reach a stationary state. We shall express the evolution time in terms of Monte Carlo sweeps. One sweep corresponds to $n$ Metropolis--Hasting updates, where $n$ is the number of nodes. The consecutive trees listed in the Markov chain  $t_0\to t_1 \to t_n \to \ldots$ are obtained one from another by one sweep. If not stated otherwise, the initial trees $t_0$ are created by either uniform or linear node attachment. In Fig. \ref{fig_slow} we show a trajectory representing  a typical evolution of the diameter of the tree during the ageing process driven
by the leaf reshuffling. 

\begin{figure}
\includegraphics[width=0.45\textwidth]{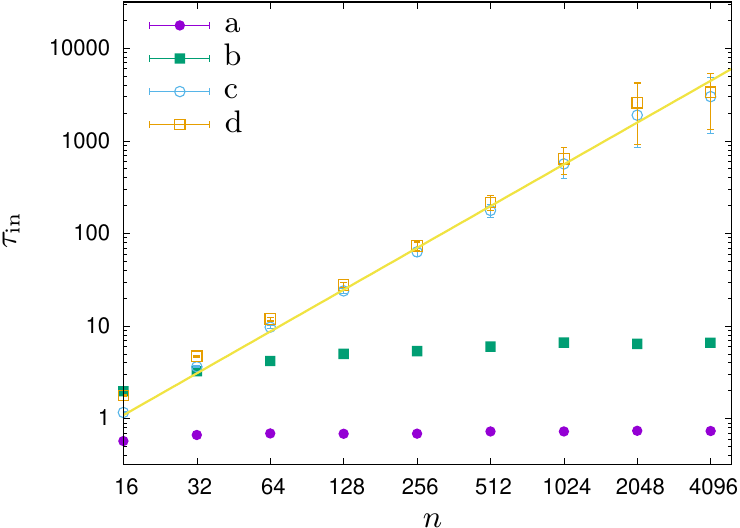}
	\caption{The autocorrelation time $\tau_{\text{in}}$ for $n=\{16,32,\ldots,4096\}$
	for the standard deviation of the node-degree distribution (a),
	for the ratio of the number of leaves to the number of nodes which are direct neighbours of leaves (b),
	for the length of the spine (c),
	and for the diameter (d).
	For the first two quantities $\tau_{\text{in}}$ increases slowly with the size and is of order ten and one, respectively for $n=4096$, while it increases rapidly for the diameter and the stem and is of order of a few thousand for $n=4094$. We plot a line corresponding to $\tau_{\text{in}}(n) = a n^z$, with $a=0.017$ and $z=3/2$ to guide the eye. 
\label{fig_tau_exp_slow}}
\end{figure}

The trajectory has typical features for ageing. It starts from an initial value and drives a long time towards an asymptotic value corresponding to the stationary state value. Once it is close to the stationary value, it begins to fluctuate around it. The consecutive values on the trajectory are correlated. The degree of correlations is measured by the integrated autocorrelation time $\tau_{\text{in}}$ which is a sort of weighted average over the wave lengths of these fluctuations. The autocorrelation time is different for different quantities. Typically one expects the autocorrelation time to asymptotically grow as a power of the system size $n$
\begin{equation}
	\tau_{\text{in}}(n) \sim n^z,
\end{equation}
when $n$ gets large \cite{Hohenberg_1977}. The exponent $z$ is sometimes called dynamic critical exponent. The autocorrelation time, and thus also the exponent $z$, depend on the dynamics of the Markov chain evolution. We have estimated values of $\tau_{\text{in}}$ for different quantities for the ageing process based on reshuffling of leaves. As an example we show in Fig. \ref{fig_tau_exp_slow} the dependence of the autocorrelation time $\tau_{\text{in}}$ for four different quantities for the exponential trees.

\begin{figure}
\includegraphics[width=0.45\textwidth]{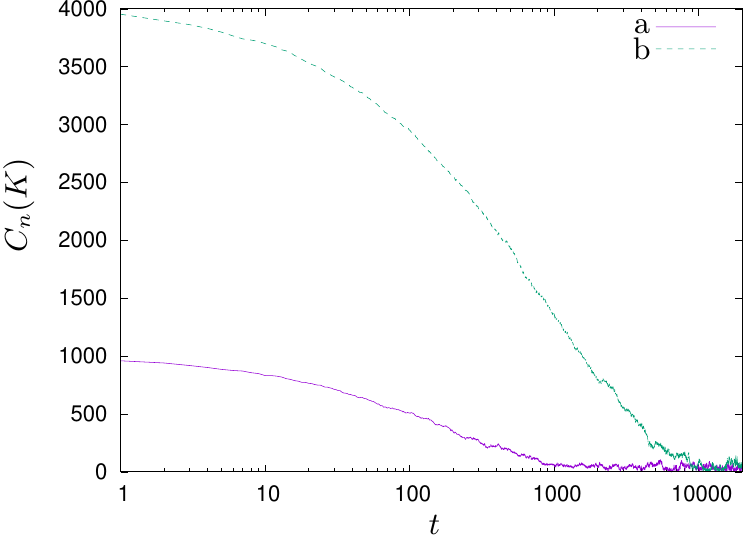}
\caption{(Colour online). We show two trajectories representing the evolution of the spine length of free trees initiated from a line graph with $n=1024$ (a) and $n=4096$ (b) nodes. The relaxation time increases with the system size. A rough estimate by unarmed eye is that the time needed to reach the stationary state value is of order one thousand in the former case and of order ten thousand in the latter one.
\label{fig_line}}
\end{figure}

We see that $\tau_{\text{in}}$ is small for local quantities related to the node degree distribution, for instance for moments of the node degree distribution. Such quantities age quickly. Also the last layers of the tree--crown age quickly. Indeed one can see from the plot that the autocorrelation time for the ratio of the number of leaves to the number of their neighbours is of order one. On the contrary, the autocorrelation time for quantities like the diameter or spine length are large and increase rapidly with $n$. These quantities are related to the global topology properties. This means that it takes a long time to rebuild the branching structure of the tree by leaf reshuffling. The reason is obvious: the process of cutting and pasting leaves operates mainly on the external layers of the tree which lie far from the spine, so it takes a long time to rebuild the spine. The spine is relatively short for the initial trees while it is long for the aged trees, as we learned in the previous section. In order to illustrate the effect we show in Fig.~\ref{fig_line} two trajectories representing a typical evolution of the spine length for initial conditions being a linear graph. As one can see from the plot, the reshuffling of leaves is very inefficient in rebuilding the spine. The trees remember the initial state for a long time and age slowly.

The problem of long range autocorrelations is a serious issue when one wants to apply the MCMC method to explore properties of the stationary state. The problem is twofold. First of all, it takes a long time to reach the stationary state, especially if the initial state lies far from it. Second of all, even if after some time the states generated by the Markov chain are close to the stationary state they may be highly correlated. This leads to an increase of statistical errors (\ref{broadening}). Intuitively, when one measures a quantity $O$ on consecutive trees $t_i$ in the Markov chain then the measurements $O_i=O(t_i)$ and $O_j=O(t_j)$ can be treated as independent only if $i$ and $j$ are separated by more than $\tau_{\text{in}}$ sweeps. This means that the MCMC sampling becomes very inefficient when $\tau_{\text{in}}$ gets large. Probably the best known example of this issue is the effect of critical slowing down known from studies of low-dimensional critical statistical systems. Near a phase transition there are usually long-range correlations between distant degrees of freedom which trigger critical fluctuations which are highly non-local. MCMC algorithms based on local update schemes are not capable to capture non-local effects properly. As a result the autocorrelation time for local algorithms is large and it quickly increases with the system size \cite{Hohenberg_1977}. An important part of the MCMC algorithm design is to reduce autocorrelations but this is a highly non-trivial task since it requires implementing non-local update schemes which are usually out of reach. A notable exception is a class of cluster algorithms applied to spin models, including the Ising model, Potts model, Heisenberg model or $O(N)$ models where whole clusters of spins are updated in a single Monte Carlo step \cite{Swendsen_1987,Wolff_1989}. They significantly reduce critical slowing down as compared to local algorithms where single spins are updated one by one. Coming back to trees, we learned that reshuffling of leaves is inefficient because is does not penetrate deeper layers of the tree--crown. An algorithm which could globally rearrange the tree structure in a single step would be more efficient. We propose such an algorithm. Instead of leaves it reshuffles whole branches of the tree. It is an adaptation of an algorithm known from Monte Carlo simulations of random triangulations and simplicial quantum gravity where it is called baby universe surgery \cite{Ambjorn_1994}. 

\begin{figure}
\includegraphics[width=0.45\textwidth]{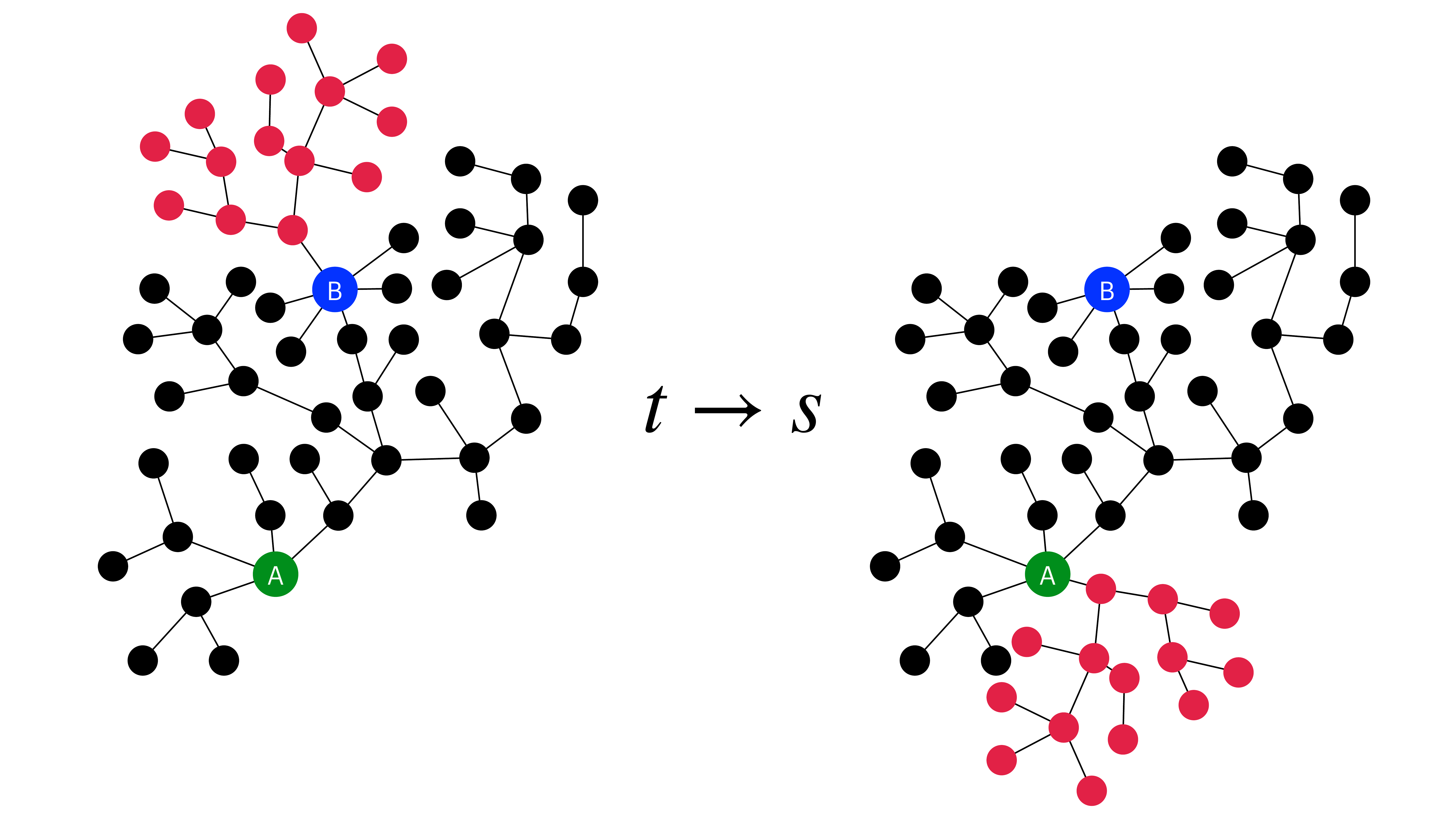}
\caption{(Colour online). An elementary update step of branch reshuffling. A branch (red) is cut from the vertex it is attached to (marked as blue B) and pasted to a randomly selected vertex (green A). As a result, the degree of the blue vertex decreases by one and of the green vertex increases by one. This transformation is analogous to the elementary transformation in the leaf reshuffling algorithm (see Fig. \ref{fig_move_leaf}) but now a much larger portion of the graph is moved from place to place in a single step.
\label{fig_move_branch}}
\end{figure}

\begin{figure}
\includegraphics[width=0.45\textwidth]{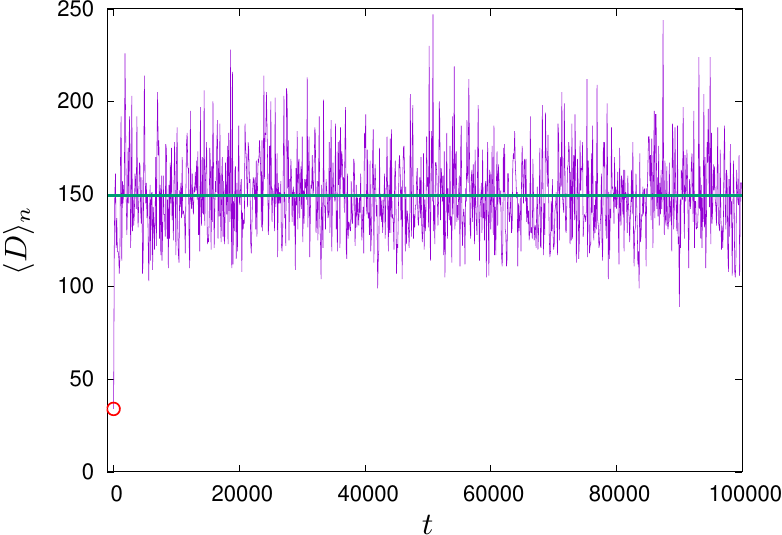}
\caption{The trajectory represents evolution of the diameter of the 
exponential tree initiated from an exponential growing tree for $n=4096$. 
The data points correspond to measurements done every 100th sweep.  The initial value, $34$, is marked by the circle and the stationary state value, $149.230(76)$, by the horizontal line. One can see that waves of oscillations around the stationary value are much shorter than in Fig. \ref{fig_slow}. 
\label{fig_fast}}
\end{figure}

In a single step of the algorithm a branch is moved from place to place, as shown schematically in Fig.~\ref{fig_move_branch}. 
The move is accepted with the Metropolis--Hasting probability which is exactly the same as for the leaf reshuffling. For weighted trees (\ref{weighted}) it depends only on degrees of the nodes between which the branch is moved. The branch is selected by choosing a random edge. There are two subtrees which grow from the endpoints of this edge. The smaller of the two, including the edge itself, is identified as the branch which is then cut and moved. The complexity of this algorithm is larger than for the leaf reshuffling because one has to make sure that the node, to which the branch is to be pasted, does not lie on the branch itself. The only way of checking this is to list all nodes on the branch. This can be done by the depth first search or breadth first search algorithms. The problem is that one does not know a priori which of the two subtrees is smaller, so sometimes it happens that one applies the search to the larger one. The worst case is when the branch is very small $n_b\ll n$ since then one may happen to explore the remaining part which has $n-n_b$ nodes. On average this strategy requires visiting half of the nodes. Since one has to do this each time when one wants to move a branch this increases the complexity of the algorithm by an extra factor proportional to $n/2$. One can however significantly reduce this factor by running the search on both sides of the edge simultaneously and stop it once all nodes on either side have been visited. In this case, instead of visiting $n/2$ nodes, one visits $2\langle n_b\rangle_n$ on average, where $\langle n_b\rangle_n$ is the mean branch size. The point is that the branch size distribution is peaked at small $n_b$ (\ref{BS}) and thus the mean branch size is much smaller than $n/2$. For example, for $n=16384$ for the scale-free trees $\langle n_b \rangle_n \approx 15$. As a result, the computer time needed for a sweep of branch reshuffling is comparable to that of leaf reshuffling while the reduction of the autocorrelation time is enormous. As an example we show a trajectory representing the evolution of the tree diameter under reshuffling of
branches in Fig. \ref{fig_fast}. 

\begin{figure}
\includegraphics[width=0.45\textwidth]{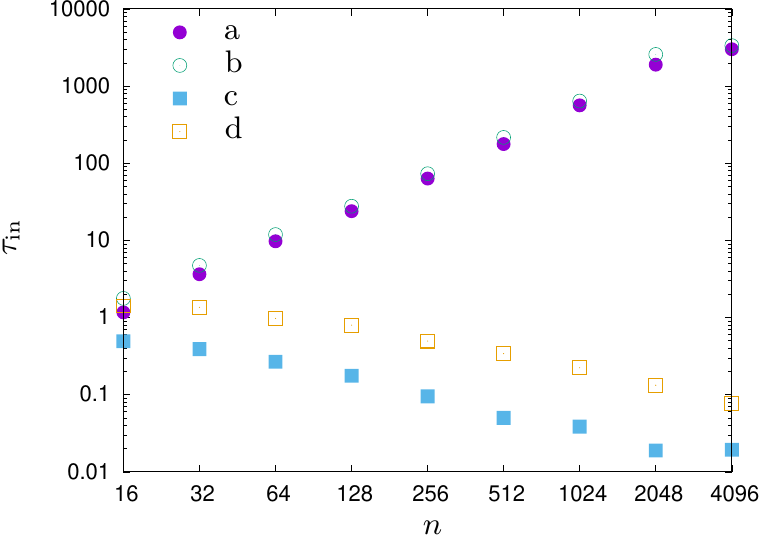}
\caption{Comparison of the autocorrelation time for the MCMC evolution of (b, d) the diameter and (a, c) the spine length of exponential trees under the leaf reshuffling (a, b) and branch reshuffling (c, d). Paradoxically the autocorrelation time for branch reshuffling decreases to zero when the system size increases. 
\label{fig_fast_slow}}
\end{figure}

It should be compared to Fig. \ref{fig_slow} where the evolution of the same system is shown but under the leaf reshuffling. For the branch reshuffling the diameter fluctuates much faster. The autocorrelations are much shorter. Also the initial part of the evolution which brings the diameter from the initial value to its stationary value lasts a few orders of magnitude shorter than in Fig. \ref{fig_slow}. 
In other words, the branch reshuffling is far more efficient as a tool to explore statistical properties of weighted tree ensembles.
To conclude this section we compare integrated autocorrelation times for leaves and branch reshuffling. As one can see in the Fig. \ref{fig_fast_slow} autocorrelations for the diameter and the spine length are close to zero when one applies the branch reshuffling algorithm.

\section{Conclusions}

We have studied ageing of trees as an example of ageing of complex networks. The initial trees were generated by a repetitive process of node-attachment. Once the trees reached a given size the growing was stopped and the further evolution followed a stochastic process preserving the size of the tree and the limiting node degree distribution. Trees evolve from the initial state to a stationary state that corresponds to maximal entropy random trees conditioned to the node degree distribution of the initial trees. We analysed the class of exponential and scale-free trees which are generated by uniform and preferential attachment rules. The statistical properties and the architecture of aged trees significantly differs from the initial ones. In particular, the diameter and the spine of initial trees increases as $\log n$ while of the aged ones as $\sqrt{n}$. Also the exponent of the branch size distribution exponent (\ref{BS}) changes from $\beta=3/2$ to $\beta=2$ which means that typical branches of aged trees are longer. The initial 
and aged trees have also completely different shapes of the tree crown distribution. The spine of the aged trees is much longer than the spine of the initial growing trees.
We have also studied relaxation properties of the ageing process for local and non-local dynamics. The local dynamics, based on leaf reshuffling is very slow in contrast to the non-local one which is based on reshuffling of branches. The typical autocorrelation time for the former one increases as a second power of the system size. 

{
In this work we have focused on trees but one can apply similar ideas to complex networks and other graph ensembles like dynamical triangulations and planar maps \cite{David_1985} or simplicial complexes which are used in simulations of quantum gravity 
\cite{Agishtein_1992,Ambjorn_1995}. When you think of complex networks, you can imagine a network that initially grows to a certain size, for example based on a preferential attachment rule \cite{Barabasi_1999} and then it ages. There are of course various ageing processes one can apply. For example the networks can be thermalised by Sneppen--Maslov transformations \cite{Maslov_2002}, or edge-rewiring transformations
\cite{Watts_1998,Burda_2001}, or a sequence of transformations which add or remove an edge with a certain probability given by a detailed-balance. In all the cases the system will relax to some equilibrium: in the first case the thermalisation will preserve the initial node-degree sequence, in the second case it will preserve the number of edges, while in the third case the number of edges will fluctuate around some specific value if the process is balanced properly. Which version of the thermalisation process is applied is a matter of the question one wants to address. With a slight abuse of terminology one can say that the three versions correspond to thermalisation in micro-canonical, canonical and grand-canonical ensembles. In all the three cases the processes are local. In general it is difficult to invent non-local transformations which would thermalise the system. The baby universe surgery that we mentioned before is rather exceptional \cite{Ambjorn_1994}. It was applied in simulations of dynamical triangulations but also in a branched polymer phase of four dimensional quantum gravity \cite{Ambjorn_1995}. There are also other non-local methods based on exact enumeration equations but they apply only to very specific models \cite{Agishtein_1992,Kawamoto_1992}. So the question if one can invent a non-local controlable dynamics for generic random graph ensembles is open. 
}

Last but not least, one could extend the studies to exotic trees obtained by tuning of the weights \cite{Bialas_1996}. In this case one can trigger the effect of the occurrence of a singular vertex of degree which is proportional to the total number of nodes. The effect is similar to the Bose-Einstein condensation. At the condensation networks undergo an interesting phase transition \cite{Bialas_1996, Bianconi_2001}.

\begin{acknowledgments}
This work was partially financed by PL-Grid infrastructure.
\end{acknowledgments}

\appendix
\section{\label{app}Trees and backgammon}

Consider a set of trees on $n$ nodes $v_1,\ldots,v_n$. The 
number of trees such that the degree of $v_1$ is $q_1$, the degree of $v_2$ is $q_2$,
{\em etc.} is given by \cite{Bryant_2015}
\begin{equation}
\frac{(n-2)!}{k_1! k_2! \ldots k_n!},
\end{equation}
where $k_i = q_i-1$, for $i=1,\ldots,n$, are nonnegative integers 
such that $k_1+k_2+\ldots +k_n = n-2$. Using this enumeration formula one can replace the sum over trees in the definition of the partition function (\ref{weighted}) by a sum over $k_i's$
\begin{equation}
\bar{Z}_{n,m} = \sum_{k_1=0}^\infty \ldots \sum_{k_n=0}^\infty 
\omega_{k_1} \ldots \omega_{k_n} \delta_{k_1+k_2+\ldots + k_n,m},
\label{balls_in_boxes}
\end{equation}
where $m=n-2$ and
\begin{equation}
\omega_k = \frac{w(k+1)}{k!}
\label{ow}
\end{equation}
for $k=0,1,\ldots$. Alternatively one can write $\omega_{q+1} = w(q)/(q-1)!$, if one replaces $k$ by $q$ in the last equation. The denominator explains the origin of the factor $(q-1)!$ discussed in main text (\ref{w_pi}). The partition function $\bar{Z}_{n,m}$ is a partition of the balls-in-boxes model \cite{Bialas_1996}, 
called also backgammon model. The backgammon model was originally proposed as a
model of entropy barriers \cite{Franz_1997,Burda_2009}. It was also used as a model of a 
real-space condensation \cite{Bialas_1997,Godreche_2005} and of zero-range processes \cite{Evans_2005}. 
Here we use it as a convenient way of enumerating trees. The model describes a statistical system of weighted partitions of $m$ particles distributed in $n$ boxes. 
The exact relation between the partition function $Z_n$ for trees (\ref{weighted}) 
and the partition function of the balls-in-boxes model is  
\begin{equation}
Z_n = (n-2)! \bar{Z}_{n,m=n-2}.
\label{zzbar}
\end{equation}
The factor $(n-2)!$ is constant for fixed $n$ and can be skipped, when $n$ is constant.
The partition function $\bar{Z}_{n,m}$ is easy to handle both numerically and
analytically. For example, it can be evaluated for finite $n,m$ by using the 
following iterative relation
\begin{equation}
\bar{Z}_{n,m} = \sum_{k=0}^m \omega_k \bar{Z}_{n-1,m-k}
\label{iterateZ}
\end{equation}
with the initial condition $\bar{Z}_{1,m} =\omega_m$. This relation immediately follows from the definition of the partition function (\ref{balls_in_boxes}).
In some particular cases a closed form solution can be given for $\bar{Z}_{n,m}$ (\ref{balls_in_boxes}). For example if the weights are $\omega_k=1/k!$ then one can easily find that
\begin{equation}
\bar{Z}_{n,m} = \frac{n^m}{m!}.
\label{free_omega}
\end{equation}
If one applies it to trees (\ref{zzbar}) one gets $Z_n=n^{n-2}$.
The asymptotic behaviour of $\bar{Z}_{n,m}$ for $n,m \to \infty$ and
$m/n \to \rho > 0 $ can be determined analytically by the saddle-point method
\cite{Bialas_1997}
\begin{equation}
\lim \frac{1}{n}\ln \bar{Z}_{n,m} = \rho \mu + f(\mu),
\end{equation}
where $f(\mu)= \ln \sum_{k=0}^{\infty} \omega_k e^{-\mu k}$ and $\mu$ is given by the equation $\rho + f'(\mu)=0$. This asymptotic formula means that the partition increases exponentially with $n$, when the number of boxes increases and the limiting density of particles approaches a constant $m/n \to \rho>0$:
\begin{equation}
\bar{Z}_{n,m} \sim e^{ m \mu + n f(\mu)}.
\label{asymptotic}
\end{equation}
For trees we have $m=n-2$ so the density is $\rho=\langle k\rangle =(n-2)/n \to 1$. The corresponding mean node degree is $\langle q \rangle  =\langle k \rangle + 1 = (2n-2)/n \to 2$ in accordance with the handshaking lemma. 

\begin{figure}
\includegraphics[width=0.45\textwidth]{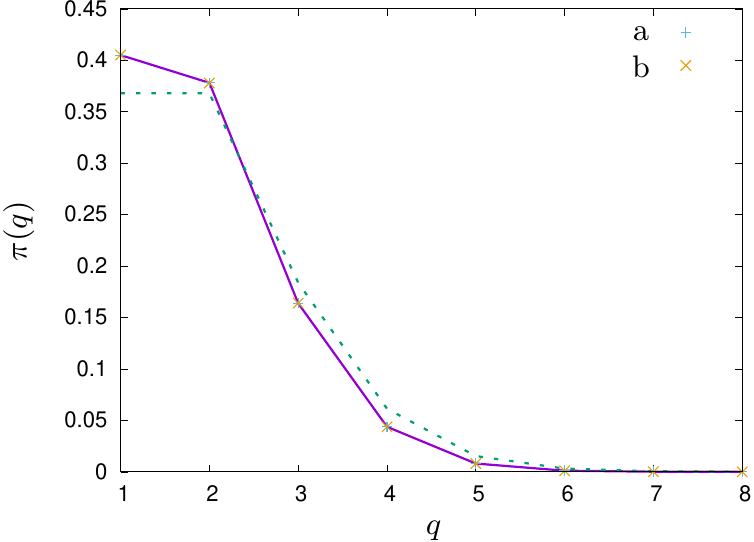}
	\caption{(Colour online) Green dotted line represents the limiting node degree distribution for free trees \eqref{pi_cayley} and purple solid line---the finite-$n$ node degree distribution \eqref{pin} for $n=16$. The lines are drawn to guide the eye.
	Symbols represent results of Monte Carlo simulations by branch reshuffling (a) and by Pr\"ufer code sampling (b) for $n=16$.
	The symbols lie on top of each other and are in a perfect agreement with the finite size prediction \eqref{pin}.
\label{fig_pi16}}
\end{figure}

In a similar way one can use the balls-in-boxes model to calculate the node degree distribution for weighted trees. The corresponding quantity in the balls-in-boxes model
is the  box-occupation probability which is defined as the probability that a box contains $k$ particles \cite{Bialas_1997,Bialas_2000}:
\begin{equation}
\bar{\pi}_{n,m}(k) = \langle \delta_{k_1k} \rangle_{n,m}=
\frac{\omega_k \bar{Z}_{n-1,m-k}}{\bar{Z}_{n,m}}.
\label{finite_size}
\end{equation} 
This is the occupation probability for the box $1$, but since all boxes are identical the occupation probability is the same for any box. The right-hand side of this equation has a clear meaning. If the box has $q$ particles, the remaining boxes form a system on $n-1$ boxes with $m-q$ particles. This formula can be applied to compute the box-occupation probability for finite $n,m$ using the iterative relation (\ref{iterateZ}).
In some cases, for example for $w(k)=1/k!$ which correspond to free trees, we can use the explicit expression for $\bar{Z}_{n,m}$ (\ref{free_omega}):
\begin{equation}
\bar{\pi}_{n,m}(k) = \binom{m}{k} \frac{(n-1)^{m-k}}{n^m}.
\end{equation} 
The corresponding node degree distribution for trees is $\pi_n(q) = \bar{\pi}_{n,n-2}(q-1)$. This gives
\begin{equation}
\pi_n(q) = \binom{n-2}{q-1} \frac{(n-1)^{n-1-q}}{{n}^{n-2}}.
\label{pin}
\end{equation}
For $n\to \infty$ the last formula approaches the limiting distribution \eqref{pi_cayley} that we discussed in the main text, but for finite $n$ it gives an exact form of node-degree distribution of free trees. We compare it with Monte Carlo data in Fig.~\ref{fig_pi16} which is, as we can see, very consistent with the theoretical finite size expression (\ref{pin}).

Using the expression (\ref{finite_size}) one can analogously compute a finite size node degree distribution for any other ensemble of weighted trees (\ref{weighted}). Inserting the asymptotic expression (\ref{asymptotic}) to (\ref{finite_size}) we can also find the limiting node degree distribution for weighted trees \cite{Bialas_1997} 
\begin{equation}
\pi(q) = \frac{w(q)}{(q-1)!} e^{-f(\mu)-q \mu}.
\end{equation}
Setting $\alpha=e^{-f(\mu)}$ and $\beta=e^{-\mu}$ we obtain the expression 
(\ref{alpha_beta}) given in the main text.

We discuss now the branch-size distribution (\ref{BS}). 
Let us first introduce the grand-canonical partition function for trees of variable size
\begin{equation}
Z(\mu) = \sum_{n=1}^\infty \frac{Z_n}{n!} e^{-\mu n} =
\sum_{n=1}^\infty z_n e^{-\mu n}.
\label{gc_pf}
\end{equation} 
The chemical potential $\mu$ is a conjugate variable to the number of vertices. The factor $n!$ is the standard symmetry factor that compensates for different permutations of vertex labels. The permutations do not change the shape. As a result, the grand canonical partition function is a sum over unlabelled trees. More precisely, the statistical weight of an unlabelled tree is inversely proportional to the volume of the automorphism group of the tree. It is worth mentioning that this type of definition is commonly used in other graph models including those in quantum gravity \cite{David_1985}, non-perturbative strings \cite{Kazakov_1985} or enumeration of Feynman diagrams \cite{Brezin_1978,Bessis_1980}.  

When $n$ is fixed the difference between $Z_n$ and $z_n=Z_n/n!$ is negligible since $n!$ is constant, however if $n$ is variable it is $z_n$ which provides a proper way of enumerating unlabelled trees. For free trees we have
\begin{equation}
z_n = \frac{n^{n-2}}{n!} \sim \frac{1}{\sqrt{2\pi}} e^n n^{-5/2},
\end{equation}
where we used the Stirling's formula to derive the large asymptotic behaviour.
Generally for weighted trees we expect that the partition function $z_n$ increases asymptotically as
\begin{equation}
z_n \sim e^{\mu_{cr} n} n^{-\gamma}
\label{zn_ab}
\end{equation}
for large $n$. The leading term is exponential. The parameter $\mu_{cr}$ corresponds to the critical value of the chemical potential, $\gamma$ is an entropy exponent which controls sub-leading corrections to the exponential growth:
\begin{equation}
\frac{\ln z_n}{n} = \mu_{cr}- \gamma \frac{\ln n}{n} + \ldots
\end{equation}
Inserting the asymptotic expression (\ref{zn_ab}) to $Z(\mu)$ 
(\ref{gc_pf}) one can see that the singular part of the grand canonical partition function 
behaves as $\sim (\mu- \mu_{cr})^{(\gamma-1)}$ for $\mu \to \mu_{cr}^{+}$. In particular for free trees (\ref{cayley}) the singular part of $Z(\mu)$ behaves as $\sim (\mu-1)^{3/2}$, which means that the second derivative of $Z(\mu)$ diverges: $Z''(\mu) \sim (\mu -1)^{-1/2}$ when $\mu \to 1^+$. The exponent $\gamma$ is universal in the sense that it is equal $\gamma=5/2$ for a broad class of generic weighted trees (\ref{weighted}). One can change it only by a very specific fine-tuning of weights \cite{Bialas_1996}. There is a very interesting way of determining $\gamma$ from a sample of trees on $n$ vertices. It is based on the observation that when one cuts a link of a tree, the tree splits into two parts being rooted trees: one with $n_b$ vertices and the other one with $n-n_b$. Denote the number of rooted trees on $n$ vertices as  $z_n'$. The number of rooted trees is related to the number of all trees as $z'_n = n z_n$ because the root can be placed at any of $n$ vertices of the tree. The branch size distribution can be calculated from the distributions of the rooted trees on both sides of the cut edge. This yields
\begin{equation}
B(n_b) \sim \frac{z'_{n_b} z'_{n-n_b}}{z_n} \sim n_b^{-\gamma+1} (n-n_b)^{-\gamma+1}.
\end{equation}
We skipped an irrelevant normalisation factor in the last formula. It can be written it the form (\ref{BS}) discussed in the main text {for $\beta=\gamma-1$. We have $B(n_b) \sim n_b^{-\beta}$} for $1 \ll n_b \ll n$ and thus this formula can be used to determine the value of the exponent. To be more precise, the two trees obtained by cutting an edge of the tree belong to a class of planted rooted trees rather than rooted trees, but the number of planted rooted trees has for large $n$ the same asymptotic behaviour as for planted rooted trees. We refer the interested reader to \cite{Burda_2003} for details.

We conclude the appendix with a short comment on the application of the Pr\"ufer code to generate free trees. The Pr\"ufer code is a one-two-one map between a set of labelled trees on $n$ vertices and a set of sequences of $n-2$ integers from the range $[1,n]$. 
Given a sequence one can unambiguously reconstruct a tree and vice versa.
This observation allows one to write a simple Monte Carlo generator of free trees. One generates a sequence of $n-2$ random integers, each being uniformly distributed on the range $[1,n]$ and converts it to a tree using the Pr\"ufer construction. Since the sequences are equiprobable, so are the corresponding trees. We used this method to test the MCMC algorithm for free trees (\ref{fig_pi16}). In all cases we observe an agreement within the statistical error between quantities computed on trees generated by the Pr\"ufer code and the MCMC method.  

{The code that we have used in Monte Carlo simulations is available at \cite{link_to_kod}.}

\bibliography{trees,km}{}
\bibliographystyle{apsrev4-2-titles}

\end{document}